\documentclass[aps,prd,twocolumn,showpacs,preprintnumbers,amsmath,amssymb,nofootinbib,scrartcl]{revtex4-1}
\usepackage{graphics} 
\usepackage{graphicx,color}
\usepackage{epsfig} 
\usepackage{bm}
\usepackage{longtable}
\usepackage{bm}
\usepackage{hyperref}
\usepackage{amssymb}
\usepackage{mathrsfs}
\usepackage{amsmath}
\usepackage{lipsum}
\usepackage{float}
\usepackage{booktabs}

\begin{document}

\title{Extended $\Lambda$CDM model}

\author{W. J. C. da Silva$^{1}$}
\email{williamjouse@fisica.ufrn.br}

\author{H. S. Gimenes$^{1}$}
\email{humbertoscalco@gmail.com}

\author{R. Silva$^{1,2}$}
\email{raimundosilva@fisica.ufrn.br}

\affiliation{$^{1}$Universidade Federal do Rio Grande do Norte,
Departamento de F\'{\i}sica, Natal - RN, 59072-970, Brazil}
\affiliation{$^{2}$ Departamento de F\'{\i}sica, Universidade do Estado do Rio Grande do Norte, Mossor\'o, 59610-210, Brasil}


\date{\today}

\begin{abstract}
In this work we discuss a general approach for the dissipative dark matter considering a nonextensive bulk
viscosity and taking into account the role of generalized Friedmann equations. This generalized $\Lambda$CDM model
encompasses a flat universe with a dissipative nonextensive viscous dark matter component, following the
Eckart theory of bulk viscosity. In order to compare models and constrain cosmological parameters, we perform Bayesian analysis using one of the most
recent observations of Type Ia Supernova, baryon acoustic oscillations, and cosmic microwave background data.
\end{abstract}

\maketitle

\section{Introduction}\label{sec1}

The observable Universe is undergoing a current process of accelerated expansion being well explained through
the standard $\Lambda$CDM model. Although this model provides a
good fit to the data, there are some drawback issues which need to be investigated, e.g., the discrepancy between the
theoretical expectation value and the observational one of the
cosmological constant \cite{weinberg89}. From the observational standpoint, there is a tension associated with
the measurement of the current value of the Hubble parameter when is used the power spectrum amplitude or considered the measurements of
the matter density parameter (see, \cite{h0tension}  and the references therein for details).
These issues also have motivated alternative models in order to study the Universe. In this concern, cosmological models
have been addressed, either using extended general relativity \cite{extendedGR}, or providing dark energy models \cite{DE}.
Some thermodynamical aspects, based on the scalar-tensor extension of the $\Lambda$CDM model has also been
presented as an argument for an extended model \cite{Algoner16}.

On the other hand, an extension of the usual Boltzmann-Gibbs Theory has been proposed in order to address the so-called
complex systems \cite{tsallis-review}.
In short, the formalism considers the entropy formula as a nonextensive quantity where there is a parameter $q$
that measures the degree of nonextensivity. The Tsallis nonextensive statistics has been successfully applied
in many physical problems \cite{tsallis-review}. This formalism was applied in cosmology scenarios, for example, entropic cosmology
for a generalized black hole entropy \cite{komatsu}, black holes formation \cite{25,27} and the modified
Friedmann equations using the Verlinde theory \cite{28}. Another direct application is the connection between
dissipative processes and nonextensive statistics \cite{29,30}. The mechanism behind this connection is based on
so-called nonextensive/dissipative correspondence (NexDC). The idea of the NexDC is associated with the microscopic
description of the fluid through the Tsallis distribution function which captures strong statistical correlation among the
4-momenta of the particles \cite{31,32}. The NexDC mechanism has been implemented in cosmology
to describe a viscous dark matter \cite{30}. By assuming the cosmological
principle, dissipative processes such as shear and heat conduction are excluded, thus, in a homogeneous and isotropic
background, only bulk viscosity is allowed for cosmic fluids. In Ref. \cite{2}, the author derived the standard theory for relativistic bulk viscosity, and some years later, connection with cosmology was derived by Weinberg,
Ellis and others \cite{3,4,5,6,7}. The introduction of bulk viscosity into cosmology has been investigated
from different standpoints. For instance, cosmological models with bulk viscosity can be interpreted as an effect of creation
of particles \cite{7-1,7a} (see, e.g., \cite{8,9,10,11,12,13,14,15,16,17,18, bulk} and the references therein for
many connections between bulk viscosity and cosmology).

An issue which can be addressed, in the alternative viewpoint of the $\Lambda$CDM model, is related to
a general framework which captures the role of the microscopic statistical correlations (nonextensive effects) introduced through the extension from the Maxwell-–Boltzmann–-Juttner statistics \cite{31,32}. Here, we propose a nonextensive $\Lambda$CDM model,
we are taking into account an extension of
standard model based on the nonextensive effects under the equipartition law of energy, as well as
an interpretation of viscous dark matter through the NexDC \cite{30}.
By assuming the Universe composed of nonextensive dissipative process (bulk viscosity), the core of the model
follows of implementation of
the nonextensive effect through the Verlinde theory \cite{19,20,21}. From a dynamical standpoint, these effects will provide
a new gravitational dynamics linked to generalized Friedmann equations. The physical motivation for the
formulation
of this extended model
is associated with microscopic statistical correlations
captured by the nonextensive framework \cite{tsallis-review}. We test the observational viability of this model performing Bayesian model selection analysis using the most
recent observations of Type Ia Supernova, baryon acoustic oscillations, and cosmic microwave background data.

This paper is organized as follows. In section \ref{sec3} we deduce modified Friedmann equations introducing the
nonextensive effect through Verlinde theory. In section \ref{sec4} we present the extended $\Lambda$CDM model.
In section \ref{sec5}, using Type Ia Supernova (SN Ia), baryon acoustic oscillations (BAO) and first acoustic peak in
cosmic microwave background (CMB) data, we implement Bayesian analysis and compare our model with $\Lambda$CDM to test the viability
of the model. Finally, in section \ref{sec6}, we summarize the main results.

\section{Friedmann Equations for dissipative processes}\label{sec3}

Let us derive the extended equations governing the dynamical evolution of the Friedmann-Lemaitre-Robertson-Walker
(FLRW) universe, from the entropic force standpoint, and taking into account the nonextensive equipartition law of energy,
the Unruh temperature, and a new interpretation for the viscous fluid. Following similar arguments of Ref. \cite{21}, the FRLW metric is given by\footnote{Here we have set $c = k_B = \hbar = 1$.}

\begin{equation}\label{eq7}
	ds^{2} = -dt^2 + a^{2}(t)(dr^2 + r^{2}d\Omega^{2}),
\end{equation}
where $a(t)$ is the scale factor of the Universe. By using the results of Ref. \cite{19},
let us consider a compact spatial region $\mathcal{W}$ with a compact boundary $\mathcal{\partial W}$,
which is a sphere with physical radius $\tilde{r} = ar$. Here, the compact
boundary $\mathcal{\partial W}$ acts as the holographic screen. By holographic principle, the number of bits on
the screen is assumed as

\begin{equation}\label{eq8}
	N = \frac{A}{G},
\end{equation}
where $A$ is the area of the screen. Assuming that the temperature $T$ on the holographic screen is related to
the total energy through the nonextensive equipartition law of energy \cite{33}
\begin{equation}\label{eq4-1}
	E_q = \frac{N}{5 - 3q}T,
\end{equation}
where $N$ is the number of bits on the screen.

Furthermore, we consider as a source of the FLRW universe, a fluid with nonextensive bulk viscosity. In this regards, the momentum-energy tensor reads \cite{30}

\begin{equation}\label{eq9}
	T_{q}^{\mu\nu} = T_{q =1}^{\mu\nu} + (q-1)\Delta T^{\mu\nu},
\end{equation}
where $T_{q =1}^{\mu\nu}$ is momentum-energy tensor of perfect fluid and $\Delta T^{\mu\nu}$ is derived of the Eckart theory, being given by

\begin{equation}\label{eq10}
	\Delta T^{\mu\nu} =\Pi h^{\mu\nu}.
\end{equation}
Here, $h^{\mu \nu} = g^{\mu\nu} + u^{\mu}u^{\nu}$ is the usual projector onto the local rest
space of $u^{\mu}$ (four-velocity) and $g^{\mu\nu}$ is the metric. $\Pi$ is the bulk
viscous pressure, which depends on the bulk viscosity coefficient and the Hubble parameter, i.e. $\Pi = -3\xi_q H$ \cite{30}.
By choosing a reference frame in which the hydrodynamics four-velocity $u^{\mu}$ is unitary, $u^{\mu}u_{\mu} = -1$, and
replacing the Eq.(\ref{eq10}) into Eq.(\ref{eq9}), we obtain
	
\begin{equation}\label{eq11}
	T_{q}^{\mu\nu} = (\rho + P_{\text{eff}})u^{\mu}u^{\nu} + P_{\text{eff}}g^{\mu\nu},
\end{equation}
where $\rho $ is the energy density, $P_{\text{eff}} = p_{k} + \Pi$, where $p_k$ is the kinetic pressure
(equilibrium pressure) and $\Pi = -3(q-1)\xi H$.
By applying the covariant derivative in Eq.(\ref{eq11}) provides

\begin{equation}\label{eq12}
	\dot{\rho} + 3H(\rho + p_k) - 9H^{2}\xi_q = 0,
\end{equation}
where $\xi_{q} = (q-1)\xi$.

The acceleration for a comoving observer at $r$ (at the place of screen) is given by \cite{21},

\begin{equation}
	a_r = -\frac{d^{2}\tilde{r}}{dt^{2}} = -\ddot{a}r.
\end{equation}
This acceleration is caused by the matter in the spatial region enclosed by holographic screen. The Unruh formula relates the temperature on the screen to an acceleration. The relation should be understood as a formula for the temperature which is related to the acceleration. In this matter, the Unruh temperature is

\begin{equation}\label{14}
T = \frac{a_r}{2\pi}.
\end{equation}
From the special relativity standpoint, we use $E = \mathcal{M}$ with $\mathcal{M}$ being the active gravitational mass,
which is related to the production of the acceleration. As is well known, this is called Tolman-Komar mass, defined by

\begin{equation}\label{15}
	\mathcal{M} = 2\int_\mathcal{W}dV\Big(T^{\mu\nu} - \frac{1}{2}Tg^{\mu\nu}\Big)u_{\mu}u_{\nu}.
\end{equation}
Here, by using momentum-energy tensor, given by Eq.(\ref{eq11}), its trace and the normalization condition $u^{\mu}u_{\mu} = -1$  as well as considering that
the active gravitational mass is measured by a comoving observer, we obtain

\begin{equation}\label{20}
	\mathcal{M} = \frac{4\pi}{3}\tilde{r}^3(\rho + 3p_k + \Pi),
\end{equation}
where $\tilde{r} = a(t)r$ and $\Pi$ is the bulk
viscous pressure (bulk viscosity). Thus, from Eqs. (\ref{eq8}), (\ref{eq4-1}), (\ref{14}), (\ref{20})
and the energy-mass relation, it is possible to show that

\begin{equation}\label{27}
\frac{\ddot{a}}{a} = -\frac{4\pi}{3}\left(\frac{5 - 3q}{2}\right)G(\rho +3p_k + \Pi),
\end{equation}
This is the acceleration equation for the dynamical evolution of the FRLW universe.
Multiplying $\dot{a}a$ on both sides of Eq.(\ref{27}) and using the continuity Eq.(\ref{eq12}), we obtain
the extended Friedmann equations

\begin{equation}
H^{2} = \frac{8\pi G}{3}\rho\left(\frac{5 - 3q}{2}\right) - \frac{k}{a^2},
\end{equation}
where $k$ is an integration constant which is identified as the spatial curvature in the region $\mathcal{W}$
in the theory of general relativity. The values for curvature are the well known, $k = -1, 0, 1$, open, closed,
flat FRLW universe, respectively. Universe without nonextensivity ($q=1$), we recovered the standard Friedmann equations.

\section{Dynamics of Nonextensive Viscous Dark Matter}\label{sec4}

Following the modified Friedmann equations deduced in the previous section, let us address the main contributions
to the total momentum-energy tensor of the cosmic fluid, i.e., the baryonic matter, the cosmological constant and
the nonextensive viscous dark matter \cite{30}. As the energy
conservation for each component of the cosmic fluid is individually conserved, we obtain

\begin{equation}
	\dot{\rho}_i + 3H(\rho_i + p_i) = 0,
\end{equation}
where $i$ corresponds baryonic matter (b), radiation (r) or cosmological constant ($\Lambda$). The conservation of
nonextensive viscous dark matter component is given by

\begin{equation}
	\dot{\rho}_{dm} + 3H(\rho_{dm} + p_{dm}^{eff}) = 0,
\end{equation}
in which $\rho_{dm}$ is the energy density and the effective pressure is

\begin{equation}\label{22}
	p_{dm}^{eff} = p_k + \Pi,
\end{equation}
where $p_k$ is equilibrium pressure (for cold dark matter $p_k = 0$) and $\Pi = -3\xi_qH$ is
the pressure from the nonextensive bulk viscosity. The equation of state, Eq.(\ref{22}) is a consequence of
the nonextensive effect, where in the limit $q \rightarrow 1$, the viscous pressure becomes null \cite{30}.
The choice of bulk viscosity coefficient $\xi_q$ seems to be an important aspect for any viscous model.
As is well known, the bulk viscosity coefficient $\xi_q$ depends on the ratio between the density of viscous
dark matter fluid at any redshift and the one today \cite{16, 17,18}

\begin{equation}
	\xi_q = \xi_{q0} \Big(\frac{\rho_{dm}}{\rho_{dm0}}\Big)^{\alpha},
\end{equation}
where $\xi_{q0}$ and $\alpha$ are constants and $\rho_{dm0}$ is the density of viscous dark matter fluid today.
Note that the present viscosity is given by the parameter $\xi_{q0}$ \cite{30}. For fixing values, $\alpha = 0$ and
$\alpha = -1/2$,
the Integrated Sachs-Wolfe effect (ISW) problem of these viscous cosmologies models is alleviated \cite{16,18}.
The values for $\alpha$ above have a physical interpretation: the lower value means a constant bulk viscosity coefficient
and the upper means, the bulk viscous fluid corresponds to the total energy. We will investigate both situations,
$\alpha = 0$ and $\alpha = -1/2$, where it will be denoted by models I and II, respectively.

\begin{figure*}
	\includegraphics[width=0.45\textwidth]{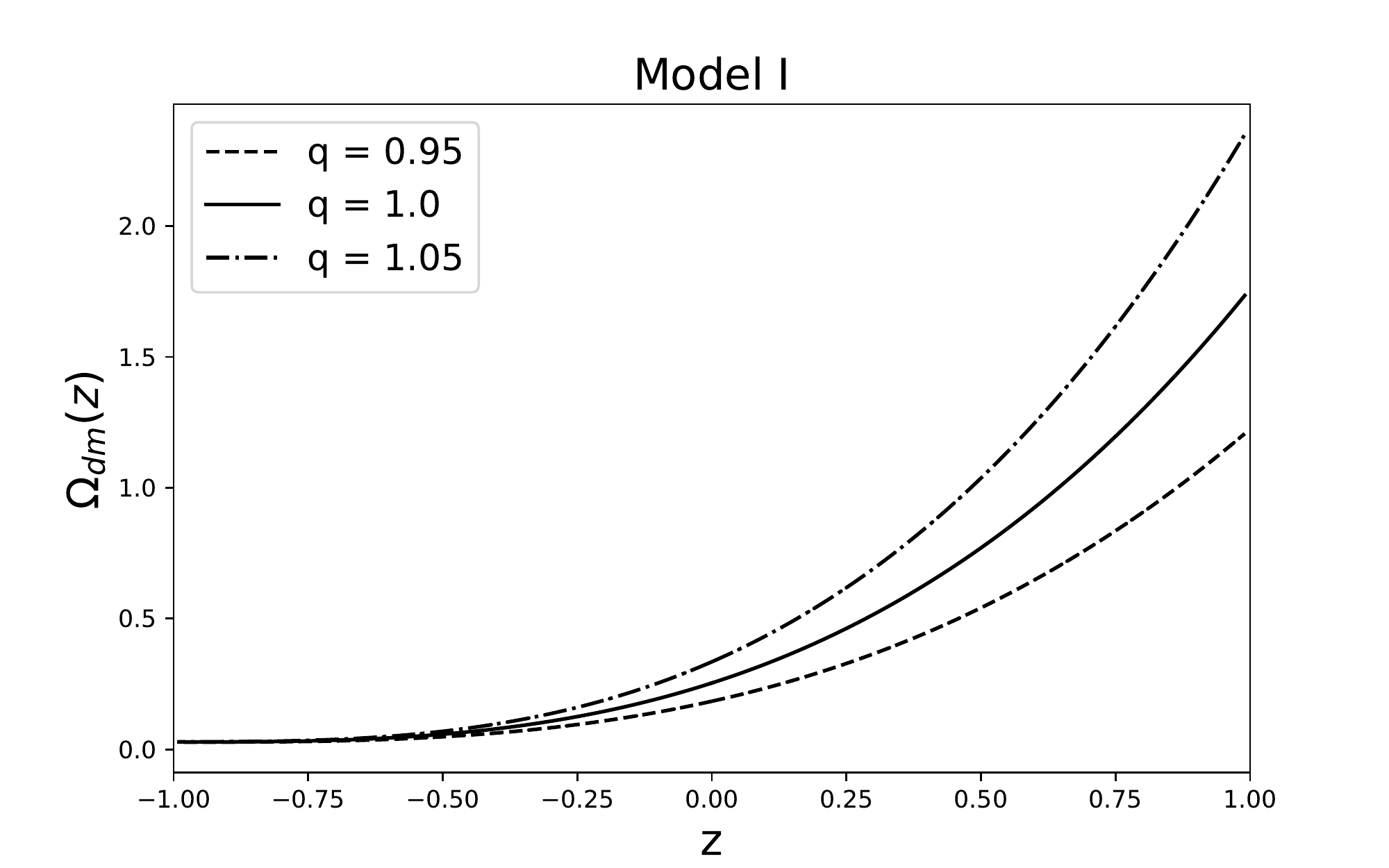}
	\includegraphics[width=0.45\textwidth]{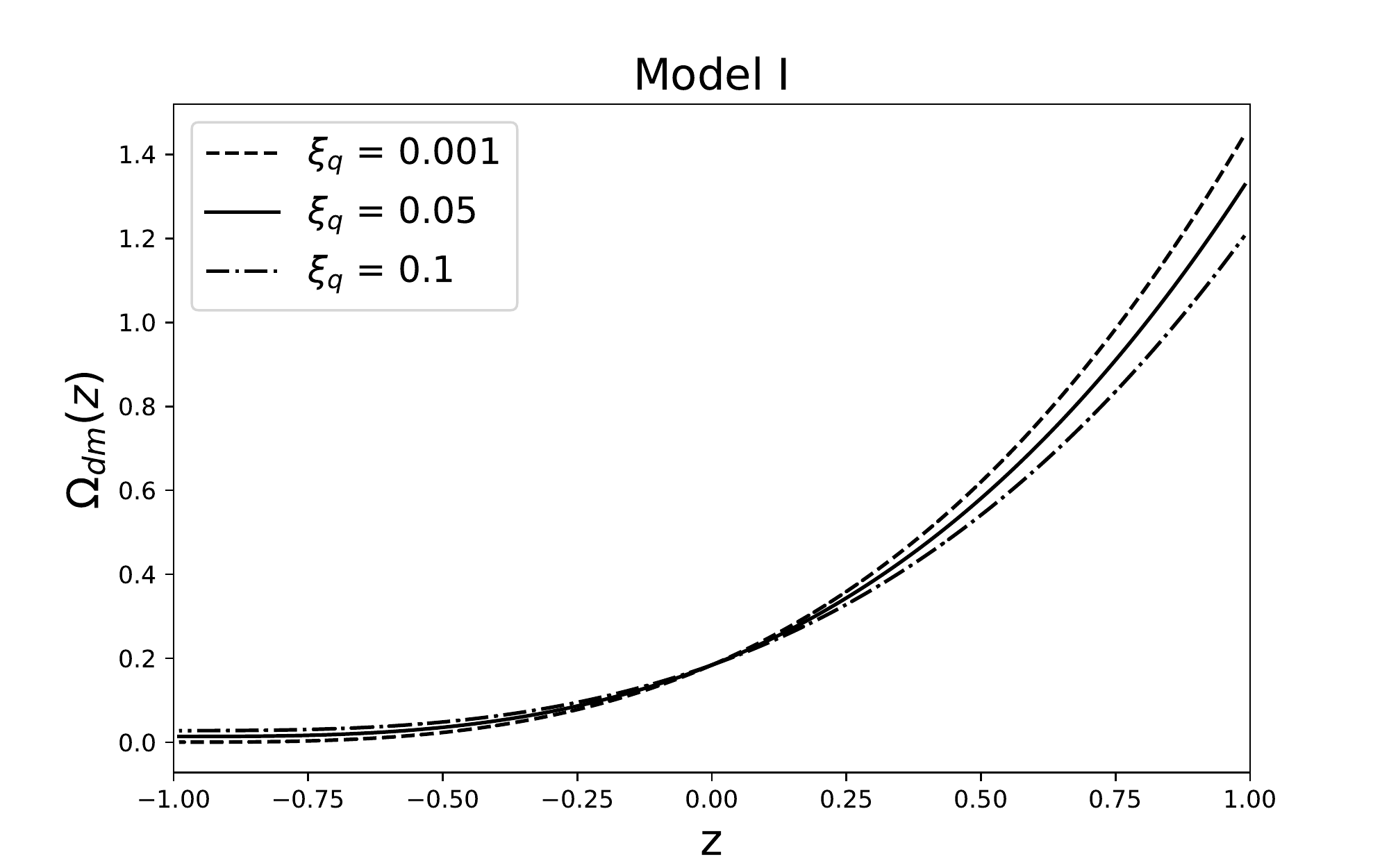}
	\includegraphics[width=0.45\textwidth]{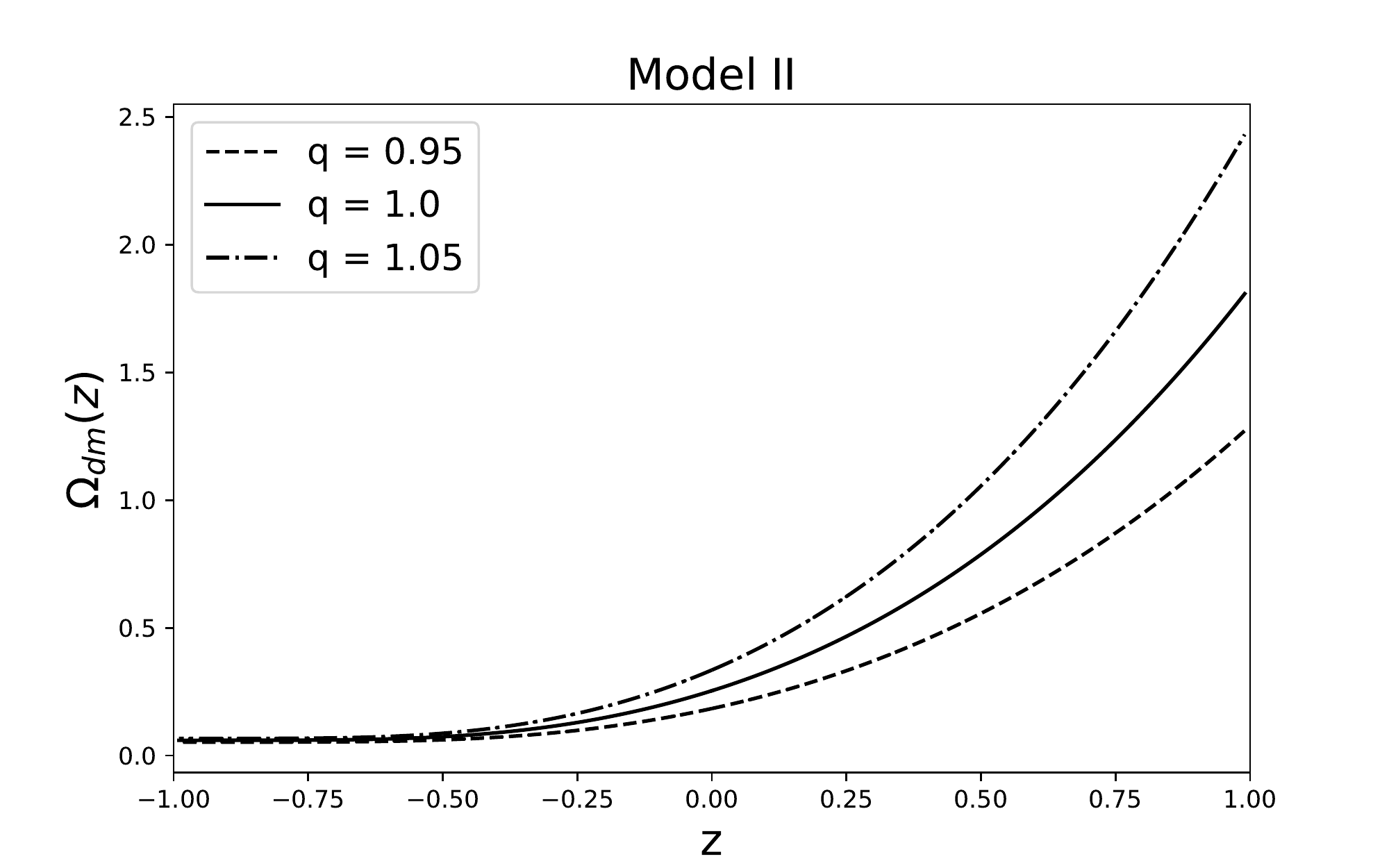}
	\includegraphics[width=0.45\textwidth]{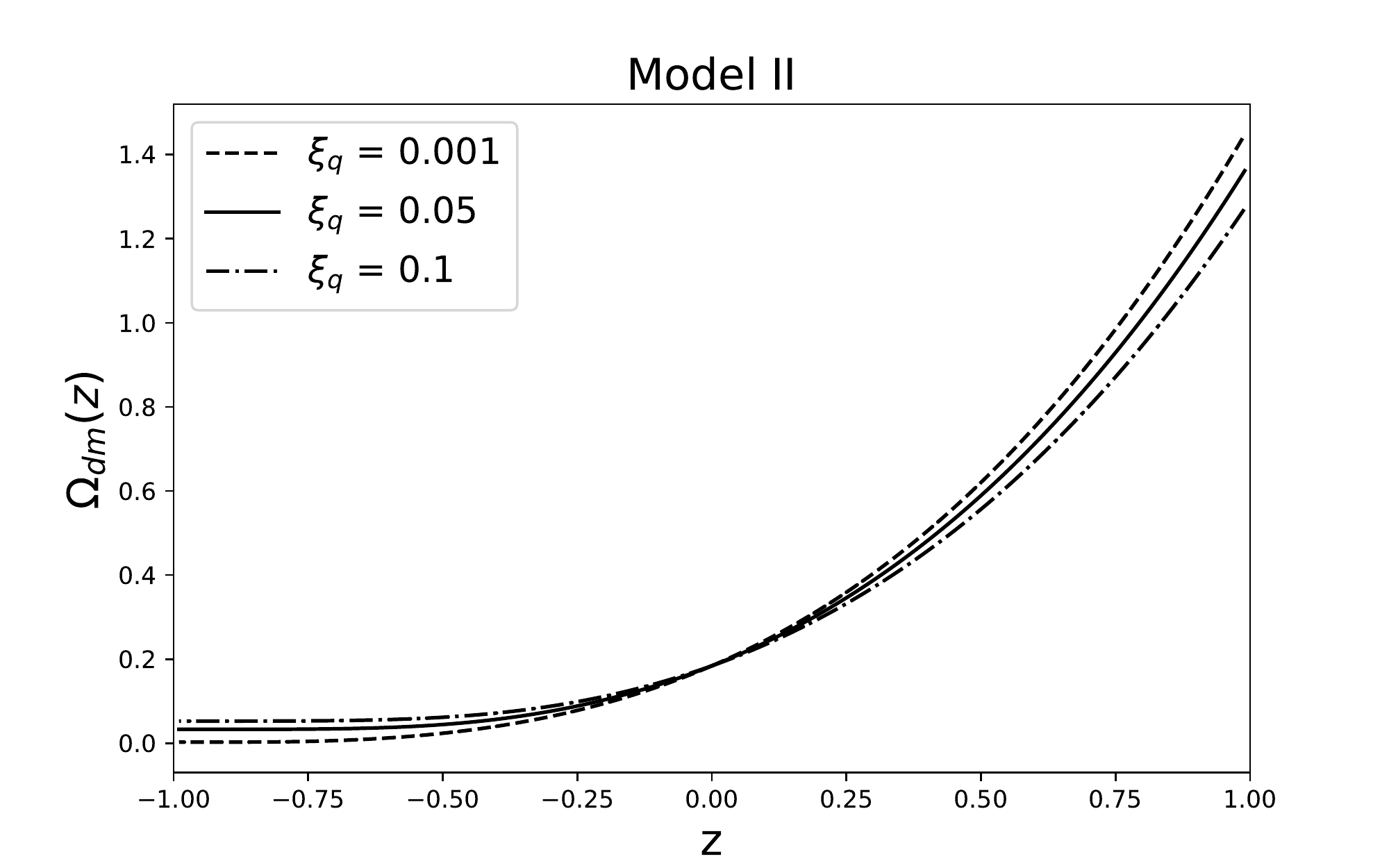}
	\caption{Evolution of nonextensive viscous dark matter parameter for selected values of $q$ and $\xi_q$.
		Here we have assumed $\Omega_{\Lambda0} = 0.70, \Omega_{b0} = 0.046, \Omega_{r0} = 8.5 \times 10^{-5}.$}
	\label{fig:evo}
\end{figure*}

The Hubble expansion rate $H$ is given in terms of the fractional energy densities $\Omega_i$, where the subscript $i$
corresponds to each fluid,

\begin{equation}
\begin{aligned}
	\frac{H^{2}(z)}{H^{2}_{0}} = \left(\frac{5 - 3q}{2}\right)\Big[\Omega_{b0}(1 + z)^{3} + \Omega_{r0}(1 + z)^{4} \\ + \Omega_{dm}(z) + \Omega_{\Lambda}\Big],
\end{aligned}
\end{equation}
In order to determine the function $\Omega_{dm}$, let us provide
the nonextensive bulk viscosity coefficient and solve its conservation equation.
For both models, the conservation equations for the nonextensive viscous dark matter fluid are given by

\begin{equation}\label{23}
\begin{aligned}
\frac{d \Omega_{dm}}{dz} = & \frac{3\Omega_{dm}}{1 + z} - \frac{\xi_q}{1 + z}
\Big[\frac{5 - 3q}{2}\Big(\Omega_{b0}(1 + z)^{3} \\ & + \Omega_{r0}(1 + z)^{4}  + \Omega_{dm} +
\Omega_{\Lambda}\Big)\Big]^{1/2},
\end{aligned}
\end{equation}
\begin{equation}\label{24}
\begin{aligned}
\frac{d  \Omega_{dm}}{dz} = & \frac{3\Omega_{dm}}{1 + z} - \frac{\xi_q}{1 + z}\Omega_{dm}^{-1/2}\Omega_{dm0}^{1/2}\Big[\frac{5 - 3q}{2}\Big(\Omega_{b0}(1 + z)^{3} \\ & + \Omega_{r0}(1 + z)^{4}  + \Omega_{dm} + \Omega_{\Lambda}\Big)\Big]^{1/2},
\end{aligned}
\end{equation}
respectively. The bulk viscosity constant reads

\begin{equation}
	\xi_q = \frac{24\pi G \xi_{q0}}{H_0},
\end{equation}
being valid for both models. The initial condition for Eqs. (\ref{23}) and  (\ref{24}) is
$\Omega_{dm}(0) = \frac{2}{5-3q} - \Omega_{b0} - \Omega_{r0} - \Omega_{\Lambda}$. The Fig. \ref{fig:evo}
shows the evolution from the nonextensive dark matter density parameter for both models considering some selected values of $q$ and
$\xi_q$. For different values of $q$ (with fixed viscosity, $\xi_q = 0.1$), both models have a similar
evolution. The models converge for a similar behavior in the future. And for different values of $\xi_q$
(with the parameter of nonextensivity fixed, $q = 0.95$), there is a small difference in the evolution at high redshifts.
It is worth noting that the nonextensivity is associated with the dynamics of the universe through the extended Friedmann
equations and with the microscopic approach for the thermodynamics of the viscous dark
matter. In particular, when $q \longrightarrow 1$, $\Lambda$CDM model is recovered. In the next section, we will use
cosmological observations in order to obtain constraints on the parameters $\xi_q$, $q$ and $\Omega_{\Lambda}$.

\section{Bayesian Analysis}\label{sec5}

Here, we will obtain the constraints of the parameters space and compare our model with $\Lambda$CDM by performing a
Bayesian statistical analysis based on the presented data. In recent years, Bayesian analysis has been widely used
to study and compare cosmological models \cite{bethoven, simony, maria, uendert, antonella}. The posterior distribution
$P(\Theta|D, M)$ is written in terms of the likelihood, $\mathcal{L}(D|\Theta, M)$, the prior, $\pi(\Theta|M)$, and Bayesian
evidence or marginal likelihood $\mathcal{E}(D|M)$ as

\begin{equation}\label{bayes}
		P(\Theta|D, M) = \frac{\mathcal{L}(D|\Theta, M) \pi(\Theta|M)}{\mathcal{E}(D|M)},
\end{equation}
where $\Theta$ denotes the parameters set, $D$ the cosmological data and $M$ the model.
The Bayesian evidence, $\mathcal{E}$, should be irrelevant in the context of the parameter estimation,
however one is essential in order to compare models based on the data.
The evidence can be written in the continuous parameter space $M$ as

\begin{equation}\label{evidence}
\mathcal{E} = \int_M \mathcal{L}(D|\Theta, M) \pi(\Theta|M) d\Theta.
\end{equation}
In order to compare two models, $M_i$ and $M_j$, both describing the same phenomenon, we compute
the ratio of the posterior probabilities, or posterior odds, given by \cite{trotta}

\begin{equation}
\frac{P(M_i|D)}{P(M_j|D)} = B_{ij}\frac{P(M_i)}{P(M_j)},
\end{equation}
where $B_{ij}$ is known as the Bayes factor, defined as

\begin{equation}\label{bayes_factor}
B_{ij} = \frac{\mathcal{E}_i}{\mathcal{E}_j}.
\end{equation}
The Bayes factor evaluates two models since a set of data, regardless of whether these models are correct.
Models with the same prior, the Bayes factor provides the posterior odds of the two models.

\begin{table}[t]
	\renewcommand{\arraystretch}{1.5}
	\renewcommand{\tabcolsep}{0.5cm}
	\centering
	\medskip
	\caption{The Jeffrey's Scale for evaluating evidence when comparing two models. The First column shows the limits
	values of the logarithm of Bayes Factor and the second column exhibits the interpretation for the strength of
	the evidence above the corresponding threshold.}
	\begin{ruledtabular}
		\begin{tabular}{cc}
			$|\ln{B_{ij}}|$& Interpretation \\ \colrule
			$ < 1$ 		   & Inconclusive   \\
			$ 1 $          & Weak           \\
			$ 2.5$         & Moderate       \\
			$ 5 $          & Strong         \\
		\end{tabular}
	\label{tab:jeffreys}
	\end{ruledtabular}
\end{table}

The Bayes factor is commonly interpreted using Jeffrey's Scale \cite{jeffreys}. In this work we use a conservative version of
 Jeffrey's scale suggested in Ref. \cite{trotta} and given in Table \ref{tab:jeffreys}. This table represents empirically
calibrated scale, with thresholds at values of $|\ln{B_{ij}}|$: $ |\ln {B_{ij}}|< 1$, the evidence in favor/against of the
model $M_{i}$ relative to model $M_{j}$ is usually interpreted as inconclusive  \cite{trotta}.
Usually the $\ln{B_{ij}} < -1$ would support model $M_{j}$. We adopt $\Lambda$CDM model as the reference model $M_{j}$.

Moreover, we consider the 1048 SNe Ia distance measurements of the Pan-STARRS (Pantheon) dataset \cite{scolnic}, the nine
estimates of the BAO parameter \cite{wigglez, bao1,bao2,bao3,bao4} and the angular scale of the sound horizon in CMB
\cite{planck} following a multivariate Gaussian likelihood given by

\begin{equation}
	\mathcal{L}(D|\Theta) \propto \exp \Bigg[-\frac{\chi^{2}(D|\Theta)}{2}\Bigg],
\end{equation}
where $\chi^{2}(D|\Theta)$ is chi-squared function for each data set.

To make this analysis, we use \textsf{PyMultiNest} \cite{johannes}, a Python interface for \textsf{MultiNest}
\cite{feroz1,feroz2,feroz3}, a generic Bayesian tool that uses nested sampling \cite{skilling} to calculate the evidence,
but which still allows for posterior inference as a consequence and we plot the results using \textsf{GetDist} \cite{42}.
Furthermore, we assume following priors on the set of cosmological parameters show in Table \ref{tab:priors}. For dimensionless
Hubble parameter $h$ we consider a range ten times wider than the value obtained in Ref. \cite{riess} and cold dark matter
parameter $\Omega_{dm}$ we use a uniform prior. For $\Omega_{\Lambda}$ we consider $68\%$ limits results of Planck 2015
\cite{planck}, we assume following prior for bulk viscosity $\xi_q$ results published in the literature \cite{16, 30}.
Moreover, for nonextensive parameter $q$ we use the limits in Ref. \cite{28}.

\begin{table}[t]
	\centering
	\medskip
	\renewcommand{\arraystretch}{1.5}
	\renewcommand{\tabcolsep}{0.2cm}
	\caption{The table shows the priors distribution used is this work.}
	\begin{ruledtabular}
		\begin{tabular}{cccc}
			Parameter      & Model & Prior 						  & Ref. \\ \colrule
			$h$			   & All   & $\mathcal{U}(0.5584,0.9064)$ & \cite{riess} \\
			$\Omega_{dm}$	   & $\Lambda$CDM   & $\mathcal{U}(0.0005,0.1)$    &  -   \\
			$\Omega_{\Lambda}$  & All   & $\mathcal{N}(0.6879, 0.0091)$   &  \cite{planck}   \\
			$ \xi_q $  & Model 1, Model 2  & $\mathcal{N}(0.0, 0.1)$   & \cite{16, 30} \\
			$q$     & Model 1, Model 2 & $\mathcal{U}(0.8, 1.10)$ & \cite{28}  \\
		\end{tabular}
	\label{tab:priors}
	\end{ruledtabular}
\end{table}

\subsection{Pantheon Supernova Type Ia Sample}

The Pantheon sample is a confirmed set of Type Ia Supernova  (SN Ia) that combine $279$ PS1 SN Ia ($0.03 < z < 0.68$) with
distance estimate of SN Ia from SDSS, SNLS, various low-z and HST samples to form the biggest combined sample of Supernova
consisting of 1048 measures ranging from $0.01 < z < 2.3$ \cite{scolnic}. By considering the instructions given in Ref. \cite{scolnic},
we use Pantheon data as if running with JLA sample \cite{37}, but the stretch-luminosity parameter $\alpha$ and the
color-luminosity parameter $\beta$ should be set to zero. So, the fundamental quantity in SN Ia analysis is the theoretical
distance modulus defined by

\begin{equation}\label{modulus}
\mu_{\text{th}} = 5 \log_{10} \frac{d_L}{\text{Mpc}} + 25,
\end{equation}
where the luminosity distance $d_L$ = $(c/H_0)D_L$, with $c$ is the speed of light, $H_0$  is the Hubble constant,

\begin{equation}\label{}
D_L = (1 + z_{\text{hel}}) \int_{0}^{z_{\text{cmb}}} \frac{dz}{E(z)},
\end{equation}
where $E(z) = H(z)/H_0$ is the dimensionless Hubble parameter, $z_{cmb}$ is the CMB frame redshift and $z_{hel}$
heliocentric redshift. In the Pantheon sample with $\alpha$ and $\beta$ equals zero, the observed distance modulus reads
\cite{scolnic, 37}

\begin{equation}
\mu_{\text{obs}} = m_\text{B} - \mathcal{M},
\end{equation}
with $m_B$ is the observed peak magnitude in rest frame B band, and $\mathcal{M}$ is a nuisance parameter that combine
absolute magnitude of a fiducial SN Ia (namely $M$) and the Hubble constant $H_0$. The $\chi^2$ function from Pantheon data
is given by

\begin{equation}
\chi^{2}_{\text{SN}} = \textbf{X}^{T}_{\text{SN}}\cdot \textbf{C}^{-1}_{\text{SN}} \cdot\textbf{X}_{\text{SN}},
\end{equation}
where $\textbf{X}_{\text{SN}} = \mu_{\text{obs}} - \mu_{\text{th}}$, and $\textbf{C}$ is the covariance matrix of
$\mu_{\text{obs}}$. It is equivalent to obtained in Ref. \cite{conley}

\begin{equation}
\chi^{2}_{\text{Pan}} = \textbf{m}^{T}\cdot \textbf{C}^{-1} \cdot \textbf{m},
\end{equation}
where $\textbf{m} = m_B - m_{\text{mod}}$, and
\begin{equation}
m_{\text{mod}} = 5\log_{10} D_L + \mathcal{M},
\end{equation}
in which $H_0$ in $d_L$ can be absorbed into $\mathcal{M}$. The total covariance matrix $\textbf{C}$ is given by \cite{scolnic}

\begin{equation}
\textbf{C} = \textbf{D}_{\text{stat}} + \textbf{C}_{\text{sys}},	
\end{equation}
where $\textbf{C}_{\text{sys}}$ and $\textbf{D}_{\text{stat}}$ are the systematic covariance matrix and diagonal
covariance matrix of the statistical uncertainty given by

\begin{equation}
\textbf{D}_{\text{stat}, ii} = \sigma^{2}_{m_{B, i}}.
\end{equation}
For \textit{i}-th SN Ia, its $m_{B, i}$, $\sigma^{2}_{m_{B, i}}$, $z_{\text{cmb}}$, $z_{\text{hel}}$ together with the
systematic covariance matrix are given by data file available in Ref.\cite{scolnic}. The nuisance parameter $\mathcal{M}$
could be marginalized following steps in Ref.\cite{conley}. We use some useful information for likelihood, therefore,
$\textbf{D}_{stat}$ and $\textbf{C}_{sys}$ are considered in this analysis.

\subsection{Baryon Acoustic Oscillations Data}
In this work, we consider an important observation to probe the expansion rate and the large-scale properties of the
universe, named baryon acoustic oscillation (BAO). The measurements of BAO provide a useful standard ruler to study the
angular-diameter distance as redshift function and the Hubble parameter evolution. The relationship between distance and
redshift can be achieved from the matter power spectrum and calibrated by CMB anisotropy data.

Commonly, the BAO measurements are shown in terms of angular scale and the redshift separation. This relation is obtained
by calculating the spherical average of the BAO scale measurement and it is given by

\begin{equation}
d_z = \frac{r_s(z_{drag})}{D_V(z)},
\end{equation}
where $D_V(z)$ is volume-averaged distance given by \cite{eisenstein, eisenstein2}

\begin{equation}
D_V(z) = \Bigg[(1+z)^2D_A(z)^2\frac{cz}{H(z)}\Bigg]^{1/3},
\end{equation}
where $c$ is the speed of light, $D_A(z) = \frac{c}{1+z}\int_{0}^{z} \frac{dz}{H(z)}$ is the angular diameter distance,
$r_s(z_{drag})$ is the comoving size of the sound horizon calculated in redshift at the drag epoch defined by

\begin{equation}
r_s(z_{drag}) = \int_{z_{drag}}^{\infty} \frac{c_s dz}{H(z)},
\end{equation}
in which $c_s(z) = \frac{c}{\sqrt{3(1 + \mathcal{R})}}$ is the sound speed of the photon-baryon fluid and
$\mathcal{R} = \frac{3}{4}\frac{\Omega_b}{\Omega_r}\frac{1}{1 + z}$. We use $z_{\text{drag}} = 1059.6$, in accordance with
Planck's 2015 \cite{planck}.

We use the BAO measurements from different surveys (see Table \ref{tab:BAO}). Additionally, we also consider three
measurements from the Wigglez survey \cite{wigglez}: $d_z(z=0.44) = 0.073$, $d_z(z=0.6) = 0.0726$, and $d_z(z=0.73) = 0.0592$. This data
is correlated by following inverse covariance matrix

\begin{equation}\label{C_BAO}
C^{-1} =
\begin{pmatrix}
1040.3 & -807.5  & 336.8 \\
-807.5 & 3720.3  & -1551.9 \\
336.8  & -1551.9 & 2914.9
\end{pmatrix} \,.
\end{equation}

\begin{figure*}[t]
	\centering
	\includegraphics[width=0.70\textwidth]{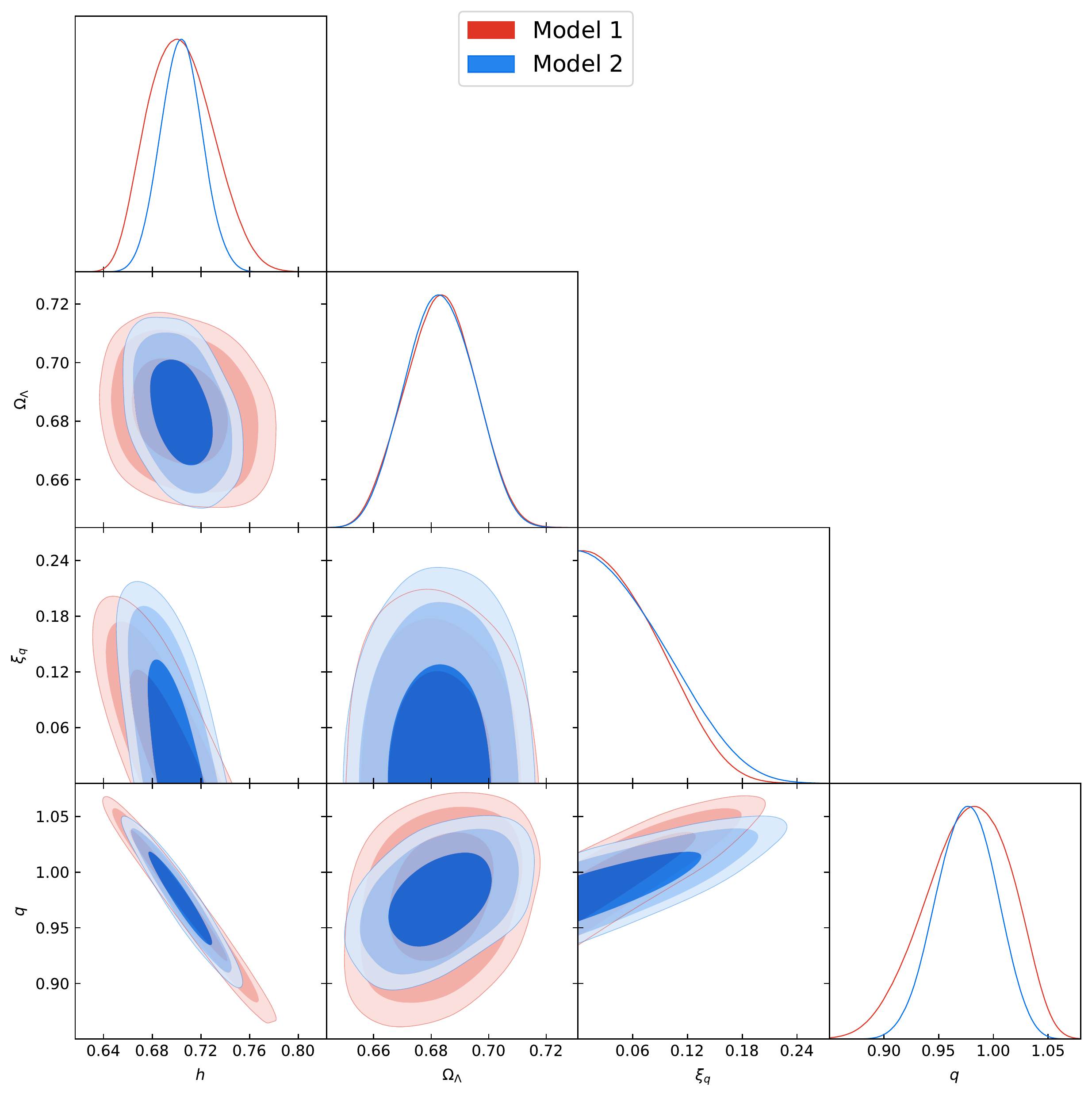}
	\caption{Confidence regions and PDFs for the parameters $h$, $\Omega_{\Lambda}$, $\xi_{q}$ and $q$, for
		the Model 1(blue) and Model 2(red) using tests with SN Ia + BAO + CMB.}
	\label{fig:regions}
\end{figure*}

\begin{table}[H]
	\centering
	\medskip
	\centering
	\renewcommand{\arraystretch}{1.5}
	\renewcommand{\tabcolsep}{0.2cm}
	\caption{\label{tab:BAO}BAO distance measurements for each survey considered.}
	\begin{ruledtabular}
		\begin{tabular}{l c c c}
			Survey     & \multicolumn{1}{c}{$z$} & \multicolumn{1}{c}{$d_z(z)$} & Ref. \\
			\colrule
			6dFGS      & $0.106$ & $0.3360 \pm 0.0150$ & \cite{bao1} \\
			MGS        & $0.15$  & $0.2239 \pm 0.0084$ & \cite{bao2} \\
			BOSS LOWZ  & $0.32$  & $0.1181 \pm 0.0024$ & \cite{bao4} \\
			SDSS(R)    & $0.35$  & $0.1126 \pm 0.0022$ & \cite{bao3} \\
			BOSS CMASS & $0.57$  & $0.0726 \pm 0.0007$ & \cite{bao4} \\
		\end{tabular}
	\end{ruledtabular}
\end{table}

For each survey considered in the Table \ref{tab:BAO}, the chi-squared function is given by

\begin{equation}
\chi_\text{Survey}^2 = \left[ \frac{d_z^{\,\text{obs}}(z) - d_z^{\,\text{mod}}(z)}{\sigma_\text{Survey}} \right]^2,
\end{equation}
where $d_{z}^{\,\text{obs}}$ is the observed ratio value, $d_{z}^{\,\text{mod}}$
is theoretical ratio value and $\sigma$ is the uncertainties in the measurements for each data point. And, for the
WiggleZ data, the chi-squared function is

\begin{equation}
\chi_\text{WiggleZ}^2 = \textbf{D}^T \textbf{C}^{-1} \textbf{D},
\end{equation}
where $ \textbf{D} = \mathbf{d}_{z}^{\,\text{obs}} - \mathbf{d}_z^{\,\text{mod}}$ and $\textbf{C}^{-1}$ is the covariance
matrix given by Eq. (\ref{C_BAO}).

Then, the BAO $\chi^2$ function contribution is defined as
\begin{equation}
\chi_\text{BAO}^2 = \chi_\text{Survey}^2 + \chi_\text{WiggleZ}^2
\end{equation}

\subsection{CMB Data}
In order to reduce the volume of the parameter space, we use the angular scale of the sound horizon at the last
scattering, defined by

\begin{equation}
\ell_a = \pi\frac{r(z_{*})}{r_{s}(z_{*})},
\end{equation}
where $r(z_{*})$ is the comoving distance of last scattering calculated in the redshift of the photon-decoupling surface,
$z_{*} = 1089.9$ \cite{planck}

\begin{equation}
	r(z_{*}) = \frac{c}{H_0}\int_0^{z_{*}}\frac{dz}{E(z)},
\end{equation}
and $r_{s}(z_*)$ is the comoving sound horizon at last scattering. We use to constraint angular scale of the sound horizon at
the last scattering data from Planck's 2015, $\ell_a = 301.63 \pm 0.15$  \cite{planck}.
The angular scale of the sound horizon at the last scattering contribution to the total $\chi^{2}$ is

\begin{equation}
	\chi^{2}_{CMB} = \frac{(\ell_{a}^{\text{obs}} - \ell_{a}^{\text{mod}})^{2}}{\sigma_{\ell_{a}}^{2}}.
\end{equation}

Therefore, the function

\begin{equation}
\chi^{2} = \chi^{2}_{\text{Pan}} + \chi^{2}_{\text{BAO}} + \chi^{2}_{\text{CMB}},
\end{equation}
which takes into account all the data sets mentioned above, should be minimized.

\section{Results and Conclusions}\label{sec6}


\begin{table*}[htb!]
	\renewcommand{\arraystretch}{2.0}
	\renewcommand{\tabcolsep}{0.2cm}
	\centering
	\medskip
	\caption{Confidence limits for the cosmological parameters using the SN Ia, BAO, CMB data.
		The first column shows the constrains on the reference ${\Lambda}$CDM model whereas the second and third
		columns show the results for the Model 1 and for the Model 2.}
	\begin{tabular}{|c|c|c|c|}
		\hline
		Parameter &${\Lambda}$CDM & Model 1 & Model 2 \\
		\hline
		\hline
		$h$ 	
		& $0.693^{+0.012+0.024+0.031}_{-0.012-0.024-0.032}$ 
		& $0.704^{+0.024+0.051+0.066}_{-0.029-0.047-0.057}$ 
		& $0.704^{+0.017+0.034+0.043}_{-0.017-0.033-0.042}$ 
		\\
		$\Omega_{dm} $	
		& $0.249^{+0.012+0.024+0.032}_{-0.012-0.022-0.028} $ 
		& $-$ 
		& $-$ 
		\\
		$\Omega_{\Lambda}$
		& $0.683^{+0.013+0.025+0.031}_{-0.013-0.025-0.031}$	
		& $0.683^{+0.012+0.023+0.027}_{-0.012-0.023-0.028}$ 
		& $0.683^{+0.012+0.022+0.028}_{-0.012-0.022-0.027}$ 
		\\
		$\xi_q$
		& $-$	
		& $0.004^{+0.074+0.14+0.16}_{-0.074-0.15-0.16}$ 
		& $-0.003^{+0.085+0.16+0.19}_{-0.085-0.16-0.19}$ 
		\\
		$q$
		& $ - $ 	
		& $0.977^{+0.043+0.069+0.082}_{-0.034-0.075-0.098} $ 
		& $0.975^{+0.027+0.052+0.062}_{-0.027-0.054-0.068} $ 
		\\
		$\ln \mathcal{E}$
		& $-529.177 \pm 0.010$
		& $-529.997 \pm 0.042$
		& $-529.745 \pm 0.017$
		\\
		$\ln B$
		& $-$
		& $ - 0.820 \pm 0.042$
		& $ - 0.568 \pm 0.017$
		\\
		Interpretation
		& $-$	
		&  Inconclusive
		&  Inconclusive
		\\
		$\xi_{q0}$
		&$-$
		&$6.0 \times 10^6(2\sigma)$
		&$6.8 \times 10^6(2\sigma)$
		\\
		\hline
	\end{tabular}
	\label{tab:results}
\end{table*}

We perform a Bayesian analysis of the nonextensive viscous models considering the evidence according to Jeffreys's scale
Table \ref{tab:jeffreys}. In this study we consider the priors shown in Table \ref{tab:priors} and background data such as,
type Ia supernova, baryon acoustic oscillations and angular scale of the sound horizon at the last scattering.  We consider
the physical constraint $\xi_{q} > 0$ upon both models in order to guarantee that second law of thermodynamics should not be
violated \cite{3,4}.

The main results of the analysis are shown in Table \ref{tab:results}, where we present the joint analysis SN Ia + BAO + CMB
with 68\%, 95\% and 99\% confidence levels (CL). In the Fig. \ref{fig:regions} show the confidence regions in $1\sigma$, $2\sigma$ and $3\sigma$ and the posterior distributions. Note that the results for both models are compatible with the $\Lambda$CDM
predictions ($q=1.00$ and $\xi_q=0.0$) at $1\sigma$ and with the results published by Refs. \cite{30, 16, 28}. We remark that
values of models parameters are slightly similar for the two models. For model 2 we have at $1\sigma$ limit $\xi_q < 0$, which
violates the second law of thermodynamics. The extended $\Lambda$CDM model was able to fit the cosmological data
at both $2\sigma$ and $3\sigma$ scenarios as well as recovered the standard $\Lambda$CDM model at $1\sigma$. The results obtained by our analysis constraint the value of Hubble constant, therefore we can calculate the discrepancy (or tension) between these values and Hubble constant local value \cite{riess2}. For Model 1, the tension is $1.078\sigma$ and for the Model 2, $1.329\sigma$, therefore the extended $\Lambda$CDM model alleviates the $H_0$ tension.

For the sake of comparison, we calculate Bayes' factor considering $\Lambda$CDM as the reference model. In Table
\ref{tab:results}, we show the values obtained for the logarithm of the Bayesian evidence ($\ln \mathcal{E}$), logarithm of
the  Bayes factor ($\ln \mathcal{B}$) and interpretation of evidence for each model considering the data. We note that
extended models are disfavored with inconclusive evidence with respect to the $\Lambda$CDM model.

For the models, the calculation of the nonextensive bulk viscosity parameter considering the $2\sigma$ value provides $\xi_q\sim 0.14$ (Model 1) and $\xi_q\sim 0.16$ (Model 2) or $\xi_{q0}\sim 10^6 Pa.s$  in SI unity. This results are in agreement from the one calculated in the Refs. \cite{44,45} in which have used the standard interpretation for bulk viscosity.

To summarize, the cosmological observations are compatible with the extended model proposed through the constraints over
$h$, $\Omega_{\Lambda}$, $\xi_q$ and $q$. In particular, the extensive limit $q=1$, the standard $\Lambda$CDM model is
recovered.

It is worth emphasizing that the microscopic nonextensive approach can be used to describe the dark energy in the context of
the bulk viscosity \cite{VDE}. Furthermore, this description can be tested through the cosmography \cite{61} and
the quintessence scenarios \cite{62}. These issues will appear in a forthcoming communication.


\begin{acknowledgements}
The authors thank CAPES and CNPq, Brazilian scientific support federal agencies, for financial support and High Performance Computing Center (NPAD) at UFRN for providing the computational facilities to run the simulations. W. J. C. da Silva thanks Antonella Cid for her support in the Bayesian analysis, Jailson Alcaniz for discussions and support.  W. J. C. da Silva would like thanks Observatório Nacional (ON) for accommodation during the development of this work.
\end{acknowledgements}

\end{document}